\documentclass[letterpaper, oneside, 12pt]{article}

\RequirePackage{amsthm, amsmath, amsfonts, amssymb}
\RequirePackage[authoryear]{natbib}
\RequirePackage[colorlinks,citecolor=blue,urlcolor=blue]{hyperref}
\RequirePackage{graphicx}
\usepackage{comment}
\usepackage{booktabs}
\usepackage{longtable}
\usepackage{array}
\usepackage{multirow}
\usepackage{wrapfig}
\usepackage{float}
\usepackage{colortbl}
\usepackage{pdflscape}
\usepackage{tabu}
\usepackage{threeparttable}
\usepackage{threeparttablex}
\usepackage[normalem]{ulem}
\usepackage{makecell}
\usepackage{xcolor}
\usepackage{stackengine}
\usepackage{lmodern}
\usepackage{setspace}
\usepackage[hmargin=1.0in,vmargin=1.0in]{geometry}
\usepackage[compact]{titlesec}
\titlespacing{\section}{0pt}{*0.8}{*0.8}
\titlespacing{\subsection}{0pt}{*0.8}{*0.8}
\titlespacing{\subsubsection}{0pt}{*0.8}{*0.8}

\newcommand{\bA}{ {\boldsymbol A} }

\newcommand{\bE}{ {\boldsymbol E} }

\newcommand{\bg}{ {\boldsymbol g} }

\newcommand{\bs}{ {\boldsymbol s} }

\newcommand{\bw}{ {\boldsymbol w} }

\newcommand{\bx}{ {\boldsymbol x} }

\newcommand{\bbeta}   { {\boldsymbol \beta} }

\newcommand{\bTheta}  { {\boldsymbol \Theta} }

\newcommand{\bpsi}    { {\boldsymbol \psi} }
\newcommand{\bPsi}    { {\boldsymbol \Psi} }

\newcommand{\bzero}{ {\boldsymbol 0} }

\title{Multi-object Data Integration in the Study of Primary Progressive Aphasia}

\author{Rene Gutierrez$^{\bf*}$\thanks{Rene Gutierrez, Assistant Professor, Department of Mathematical Science, UTEP, 500 W. University Ave. El Paso, TX 915-747-5000  (E-mail: rgutierrezmar@utep.edu).} \and Rajarshi Guhaniyogi$^{\bf*}$\thanks{Rajarshi Guhaniyogi,  Associate Professor, Department of Statistics, Texas A \& M University, TAMU 3143, College Station, Texas 77843-3143 (E-mail: rajguhaniyogi@tamu.edu).}
\and Aaron Scheffler$^{\bf*}$\thanks{Aaron Scheffler,  Assistant Professor, Department of Epidemiology and Biostatistics, UC San Francisco, 550 16th Street, San Francisco, CA 94158 (E-mail: Aaron.Scheffler@ucsf.edu ).}
\and Maria Luisa Gorno-Tempini$^{\bf*}$\thanks{Maria Luisa Gorno-Tempini,  Professor of Neurology and Psychiatry, Department of Neurology, UC San Francisco, UC San Francisco, 550 16th Street, San Francisco, CA 94158 (E-mail: marialuisa.gornotempini@ucsf.edu).}
\and Maria Luisa Mandelli$^{\bf*}$\thanks{Maria Luisa Mandelli,  Associate Adjunct Professor, Department of Neurology, UC San Francisco, UC San Francisco, 550 16th Street, San Francisco, CA 94158 (E-mail: MariaLuisa.Mandelli@ucsf.edu).}
\and Giovanni Battistella$^{\bf*}$\thanks{Giovanni Battistella,  Instructor in Otolaryngology–Head and Neck Surgery, Department of Otolaryngology–Head and Neck Surgery, Harvard Medical School (E-mail: gbattistella@meei.harvard.edu).}}

\begin{document}
\maketitle

\begin{abstract}
This article focuses on a multi-modal imaging data application where structural/anatomical information from gray matter (GM) and brain connectivity information in the form of a brain connectome network from functional magnetic resonance imaging (fMRI) are available for a number of subjects with different degrees of primary progressive aphasia (PPA), a neurodegenerative disorder (ND) measured through a speech rate measure on motor speech loss. The clinical/scientific goal in this study becomes the identification of brain regions of interest significantly related to the speech rate measure to gain insight into ND patterns. Viewing the brain connectome network and GM images as objects, we develop an integrated object response regression framework of network and GM images on the speech rate measure. A novel integrated prior formulation is proposed on network and structural image coefficients in order to exploit network information of the brain connectome while leveraging the interconnections among the two objects. The principled Bayesian framework allows the characterization of uncertainty in ascertaining a region being actively related to the speech rate measure. Our framework yields new insights into the relationship of brain regions associated with PPA, offering a deeper understanding of neuro-degenerative patterns of PPA. The supplementary file adds details about posterior computation and additional empirical results. 
\end{abstract}


\section{Introduction}
\noindent 
This article is motivated by a clinical application on patients suffering from brain loss due to Primary Progressive Aphasia (PPA), a neurodegenerative disorder (ND) characterized by loss of language ability that shares pathological signatures with Alzheimer’s disease and frontotemporal dementia. Despite the devastating toll PPA effects on a patient's ability to communicate, it is far less widely studied than other NDs, and as a result, research cohorts contain modest sample sizes. We consider the nonfluent/agrammatic variant of PPA (nfvPPA) characterized by motor speech and grammar loss and left inferior frontal atrophy \citep{gorno2008}. Multi-modal imaging data are available for each of these PPA patients which include: (a) \emph{brain network information} and (b) \emph{brain structural information}. Brain networks consider each region of interest (ROI) as a network node and quantifies the connectivity between the pairs of network nodes using functional magnetic resonance imaging (fMRI). Brain structural information is obtained using structural magnetic resonance imaging (sMRI), e.g., gray matter (GM) images over brain volumetric pixels (voxels). Both images are collected on a common brain atlas which segments a human brain into different ROIs, with each ROI containing a number of voxels. A speech rate score is available for the patients to measure the degree of their motor speech loss due to PPA.

The major scientific goals of the study are twofold, (a) to estimate regression relationships between the speech rate and the two imaging modalities; and (b) to identify ROIs significantly related to the speech rate, and thus infer the ND patterns for PPA.
While many ROI-based inferential techniques are constrained by considering the ROI as the most basic unit of analysis, the PPA data application allows for additional granular inference at the level of the voxel which will provide neuroscientists with the dual ability to frame results at the macro level of the ROI and probe the micro voxel-level signals that drive these results. To our knowledge, this article proposes the first statistical framework that allows for flexibility of macro and micro inference on multi-object imaging data which is urgently needed in neuroscience \citep{calhoun2016multimodal}. Uncertainty quantification regarding the inference on influential regions becomes crucial since the analysis only involves a limited number of rarely studied PPA patients. 

This article addresses the inferential objective by formulating a multi-modal regression framework with the brain network (defined over brain ROIs) and structural images (defined over brain voxels) as two sets of responses and the speech rate as the predictor. 
A hierarchical Bayesian approach is adopted wherein an integrated prior structure on coefficients of the speech rate corresponding to the structural and network objects (hereon referred to as the structural and network coefficients, respectively) is developed to allow the information in separate image objects to complement and reinforce each other in their relation to the scalar predictor. To elaborate, we begin with a common brain atlas for both image objects to ensure an organizing principle that links together structural and network information via a shared set of ROIs and a group of voxels within each ROI. The integrated prior on network and structural coefficients are constructed respecting the \emph{hierarchical} constraint which ensures all voxels within an ROI uninfluential if the ROI is un-influential (see Section~\ref{sec:prior_coef}). While we do not explicitly make use of the structural information in the GM images by careful spatial modeling of the GM image coefficient, it partially exploits the structural information by imposing the \emph{hierarchical} constraint in the prior construction. The problem of identifying influential ROIs is cast under a nonlinear variable selection framework, wherein latent activation indicators corresponding to all ROIs are shared among both sets of coefficients to enforce that all voxels from a particular ROI and all network edges connected to that particular ROI have no relationship with the predictor when the activation indicator corresponding to the ROI is zero. As a byproduct of our construction, the \emph{symmetry} and \emph{transitivity} properties \citep{hoff2005bilinear} of the un-directed network object are preserved, as discussed in Section~\ref{sec:prior_coef}. The prior construction achieves efficient computation, identifies influential ROIs and voxels within an influential ROI which are key to studying neuronal degradation, and produces well-calibrated interval estimates for the multi-modal regression coefficients. Moreover, our framework attaches uncertainty in identifying these ROIs and offers improved inference over regression methods with a single imaging modality.


There is a dearth of principled Bayesian literature addressing the inferential objectives of the motivating application, and our proposal is arguably the first Bayesian multi-object regression approach to answer the inferential questions stated before. We now provide a brief overview of the available literature to contrast them with our proposal. In the course to determine the association between network or structural objects and the speech rate score, the most popular approach estimates the association between each network edge or GM voxel and the language score independently, providing
a p-value ``map" \citep{lazar2008statistical}, after adjusting for multiple comparisons to identify ``significant" ROIs.  Such approaches have key disadvantages relative to methods that take into account the integrated impact of all network edges and voxels of the GM images simultaneously, as demonstrated in Section~\ref{sec:sim_study}. 

One can possibly proceed to vectorize both objects and regress them jointly on the speech rate with high dimensional multivariate reduced rank sparse regression \citep{chen2012sparse, goh2017bayesian}. However, 
vectorization of object responses during analysis ignores their individual characteristics (e.g., the symmetry and transitivity in the network) and their interconnections through the hierarchical constraint, and apparently does not allow identification of influential ROIs with uncertainty.

There have been some recent efforts to build regression models with a single-object response (e.g., tensors, networks, or functions), mainly for neuro-imaging data \citep{spencer2020joint, zend2021bayesian, guha2021bayesian}. While these approaches establish the importance of exploiting the structure and/or network characteristics in image objects for better inference, principled linkages among image objects are not made (e.g., through the hierarchical constraint) and thus inference is made without regard to information shared across these objects. Failure to consider the structure and cross information from images have generally a negative impact on ND research in terms of bias in estimated effects, statistical inefficiency \citep{dai2021orthogonal}, and sensitivity of results to noise \citep{calhoun2016multimodal}. 

Our proposed approach is considerably different from the existing statistical literature on multi-modal data integration. In particular, there has been a class of unsupervised multi-modal analysis with methods exploiting structural connectivity information from diffusion tensor imaging (DTI) in the prior construction for the functional connectivity analysis from functional MRI (fMRI) data \citep{xue2015multimodal}. Building on work in sparse canonical correlation analysis \citep{witten_extensions_2009, chen_structure-constrained_2013}, there is a growing literature in imaging genetics that seeks to model the cross-covariance between a set of high-dimensional genetic features and imaging phenotypes from different modalities \citep{du_fast_2018, Zhang2022} which occasionally integrates linkage disequilibrium into estimation but does not explicitly model the topology of the imaging data. With regard to supervised learning approaches, nearly all approaches treat multi-modal imaging data as a predictor rather than a response. \cite{xue2018bayesian} proposes regression of disease status on low-frequency fluctuation (fALFF) from resting-state fMRI scans, voxel-based morphometry (VBM) from T1-weighted MRI scans, and fractional anisotropy (FA) from DTI scans. In the same vein \cite{li2021integrative, dai2021orthogonal} develop frameworks to account for the non-linear association between a scalar response and multi-modal predictors. \citet{peng_structured_2016} utilizes sparse kernel learning with a novel group penalty to integrate information among genetics and multi-modal images to predict Alzheimer's status. 
\citet{Ma2023, Guo2022} both develop Bayesian regression models that predict a single response image from multi-modal image predictors. Recently, there has been a flurry of attention focused on generative deep learning which seeks to generate multi-modal images based on input features including disease status \citep{Suzuki2022, Fernandez2023} for the purpose of data augmentation and synthetic image generation. These approaches commonly employ regression frameworks with a scalar response and multi-modal imaging predictors which presents two major challenges for our application of interest. First, the limited sample size in our study (sample size of 24, see Section~\ref{PPA}) necessitates significantly downsizing each image and assuming strong sparsity in the image predictors to achieve statistically meaningful inferences from the high-dimensional regression, which is less appealing to neuroscientists. Second, image voxels often exhibit collinearity, resulting in less reliable inference when used as predictors in high-dimensional regression. We address this issue by performing integrated modeling of structural and network images as a function of a primary predictor of interest. Our framework is able to deliver inference on influential ROIs which is the main scientific objective of the clinical study. Notably, all these aforementioned approaches do not naturally offer identification of influential ROIs with uncertainty.

Up to this point, the foundational work in the cognitive neuroscience of PPA \citep{mandelli2016, gorno2011, gorno2008} has not directly combined structural information relating to neuronal atrophy with network information on brain connectivity. At best, brain networks that underlie language have been identified via an informal fusion of information from multi-object image data via seed-based methods \citep{battistella2020}. Specifically, several ROIs are identified that display the highest degree of atrophy in structural images (e.g. GM degradation) of PPA patients. These ROIs are then used to define ``seed regions" which are used to construct speech networks that quantify the nodes highly connected to the seed regions. While the speech networks identified via this approach are likely important to study PPA, the ad-hoc way of selecting seed regions means that the influence of atrophy in other ROIs may have been overlooked and inference may have been subject to the limitations of unimodal analysis. 

 The rest of the manuscript proceeds as follows. Section~\ref{PPA} 
 provides a description of the structure of the multi-modal data we analyze in this article. Section~\ref{sec:model_development} describes the model development and prior framework to draw inference from multi-modal images and Section~\ref{posterior_computation}
 discusses posterior computation of the proposed model. Empirical investigations with data generated under various simulation settings are reported in Section~\ref{sec:sim_study}. Section~\ref{PPA_analysis} analyzes the multi-modal dataset, offering scientific findings on influential ROIs. Finally, Section~\ref{Conclusion} summarizes the idea and highlights some of the extensions to be explored in the near future.

 \section{Clinical Case Study on Nonfluent Primary Progressive Aphasia (nfvPPA)}\label{PPA}
We focus on a clinical application derived from multi-modal image studies conducted on patients with PPA, an ND that impacts brain networks that support language \citep{gorno2011}. Our interest lies in the nonfluent/agrammatic variant of PPA (nfvPPA) characterized by motor speech and grammar loss and left inferior frontal atrophy \citep{gorno2008}. To investigate the neural underpinnings of disruption to motor speech/fluency in nfvPPA patients, clinical images from multiple modalities were collected as detailed below. Only a handful of institutions worldwide collect brain scans on PPA patients for research, and thus our sample size is constrained due to the prevalence of the disease and resource limitations. However, the utility of the disease as a lens to understand the neuroanatomy of language production provide sufficient clinical motivation to proceed with analysis while practicing sufficient caution in interpretation due to the modest sample size. 


\begin{figure}
\begin{tabular}{c}
\includegraphics[trim = 0in 0in 0in 0in, clip, width=0.6\linewidth]{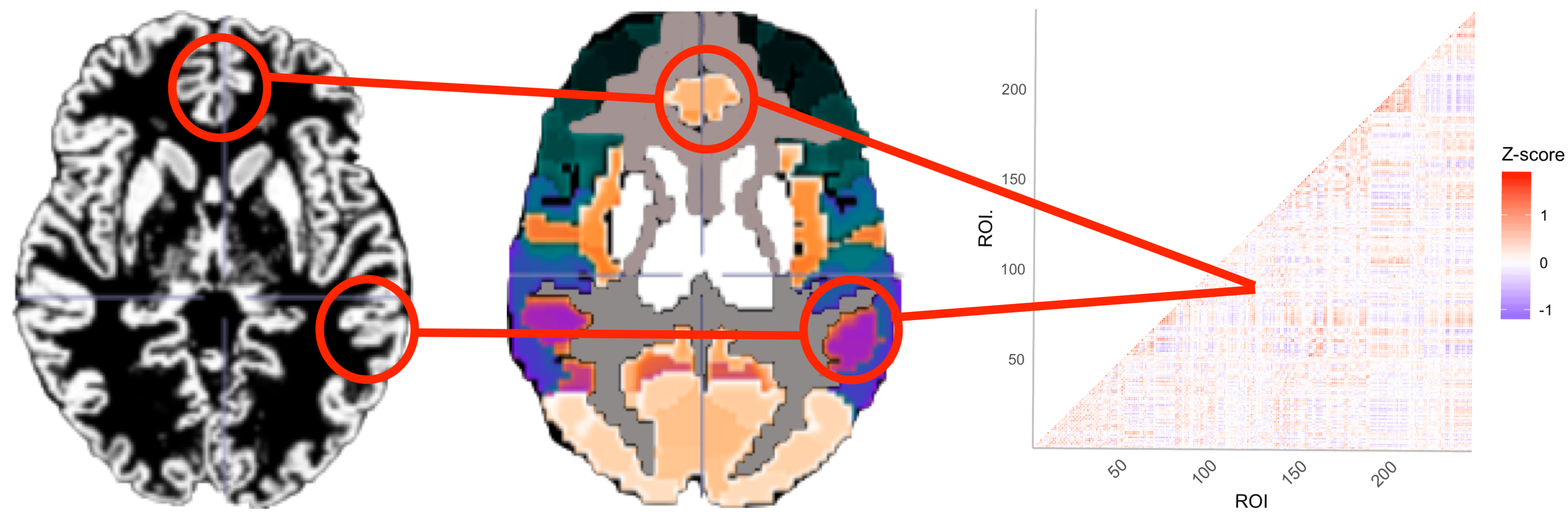}\\
\small{(a) Structural image \quad \quad \quad \quad (b) Atlas \quad \quad \quad \quad \quad \quad (c) Network image \quad \quad}  \\
\end{tabular}
\caption{\footnotesize{Schematic of the multi-object brain imaging data structure for a PPA patient. (a) Structural image encoding voxel-level gray matter (GM) probability, (b) Brainnetome atlas parcellation of the brain into anatomical ROIs, (c) Network image obtained by calculating the pairwise Pearson correlation z-score for the average fMRI signal in each ROI. Red circles and lines connect (a) structural and (c) network information from images via the (b) parcellated atlas. Thus, the atlas provides an organizing hierarchy that links together structural information (GM) at the voxel level with network information indexed by pairs of ROIs (fMRI).}}
\end{figure}

\noindent\underline{\textbf{Clinical images and language evaluation:}} PPA being an infrequently phenotyped ND, imaging data is acquired on $24$ nfvPPA patients during the course of clinical research activity. Data is collected from the following imaging modalities: sMRI derived GM values (Figure 1a) which measures the segmentation probability that a voxel contains neuronal cell bodies \citep{rajapakse}, and task-free resting state fMRI to measure brain activation via neuronal oxygen consumption in subjects at rest. All images are registered to the Montreal Neurological Institute (MNI) template space with voxels parcellated into 245 ROIs using the Brainnetome atlas such that images across modalities and subjects can be directly compared and each voxel is nested in an anatomically defined ROI {\textbf{(Figure 1b)}} \citep{brainnetome}. The choice of brain atlas is an important decision in the analysis pipeline as it impacts subsequent results and interpretations. Based on our collaborators' prior work studying language in patients with PPA, the Brainnetome atlas is selected since it provides a granular parcellation needed to model the subtle structural and network dynamics of speech fluency.   
For each subject, a `brain network' represented by a symmetric adjacency matrix is obtained from the fMRI image by considering rows and columns of this matrix corresponding to different ROIs and entries corresponding to the z-scores obtained by Fisher Z-transforming the Pearson correlation of the average fMRI data of two ROIs {\textbf{(Figure 1c)}}. We focus our analysis of language loss on speech rate, the number of words spoken per minute, a measure of motor speech via a subject’s articulation rate. This speech rate measure is automatically extracted from recorded speech from the Grandfather passage, a 129-word block of text meant to elicit a comprehensive set of phonemes in English \citep{grandfather}, via SALT software (https://www.saltsoftware.com/), a software platform used to automatically extract language features from recorded speech.
 
\noindent\underline{\textbf{Scientific question of interest:}} 
 Language loss in nfvPPA patients is driven by neurodegeneration in the \emph{left inferior frontal} region but the dual role of structural damage and brain connectivity in language loss is not well characterized \citep{mandelli2016}. Findings from the prior clinical studies based on individual image modalities have identified brain networks related to motor speech/fluency \citep{mandelli2016}. One of our primary goals is to extend the study of motor speech/fluency to consider multi-modal image data spanning the whole brain to better understand how structural and network changes \emph{simultaneously} associate with speech ability. This scientific question is motivated by findings in PPA and other NDs that observe alignment of disruptions in functional connectivity and structural atrophy \citep{Johnson2000, Battistella2019, brown, raj} and thus our inference inherently focuses identifying and quantifying these co-localized effects and their relation to language loss. The exact mechanism underlying this co-localization is not known, and may not fully explain language loss in nfvPPA patients, but characterizing and measuring these co-localized effects is only possible by analyzing multiple image modalities and is needed to advance understanding of disease progression and presentation. To do so, we apply our proposed multi-modal framework to regress sophisticated multi-modal images, specifically GM maps which capture focal neurodegeneration, and fMRI brain connectivity networks which capture disruptions of brain connectivity, on the speech rate score.
 The next section describes the novel regression framework to achieve these scientific goals.

\section{Bayesian Regression with Multiple Imaging Responses and Speech Rate as a Predictor}\label{sec:model_development}
\noindent This section presents model development and prior formulation, including the hyper-parameter specification.

\subsection{Model Framework}
\noindent For the $i$th subject, let $y_i\in\mathbb{R}$ denote the speech rate measure, $\bx_i=(x_{i,1},...,x_{i,H})'$ represent the biological and demographic variables and $\bA_i$ represent the weighted brain network object. We assume that the network of all subjects is defined on a common set of nodes, with elements of $\bA_i$ encoding the strength of network connections between different nodes for the $i$-th subject. In particular, the network object $\bA_i$ is expressed in the form of a $P\times P$ matrix with the $(p,p')$-th entry of the matrix $a_{i,(p,p')}$ signifying the strength of association between the $p$th and $p'$th node, where $p,p'=1,...,P$ and $P$ is the number of network nodes. This paper specifically focuses on networks that contain no self-relationship, i.e., $a_{i,(p,p)} \equiv 0$, and are un-directed ($a_{i,(p,p')}=a_{i,(p',p)}$). Such assumptions hold for the data application pertaining to Section~\ref{PPA}, where $\bA_i$ represents the brain connectome network matrix obtained from the fMRI scan, with each node representing a specific brain ROI and edges signifying correlations between fMRI signals in two regions. Let $\bg_{i,1}$,...,$\bg_{i,P}$ denote the $V_{1},..., V_{P}$ dimensional structural objects in regions $\mathcal{R}_1$,...,$\mathcal{R}_P$, respectively. In the context of our data application, they represent voxels of the GM image from the $P$ ROIs. 

For $i=1,\ldots,n$, we assume that the relationship between the speech rate $y_i$ varies with every network edge and every GM voxel after accounting for the edge and voxel-specific effect of the biological and demographic covariates, and propose conditionally independent generalized linear models for every network edge and GM voxel, given by
\begin{align}\label{joint_model}
E[a_{i,(p,p')}] = H_1^{-1}\Big(\sum_{h=1}^{H} \psi_{p,p',h}^a x_{i,h} +\theta_{p,p'}\: y_i\Big),\:\:
E[g_{i,v,p}] = H_2^{-1}\Big(\sum_{h=1}^{H} \psi_{p,h}^g x_{i,h} + \beta_{v,p}\: y_i\Big),
\end{align}
for $v=1,\ldots,V_p;\:\:p=1,\ldots,P$,
where $H_1(\cdot)$ and $H_2(\cdot)$ are the link functions, $\theta_{p,p'}$ is the $(p,p')$th element of the $P\times P$ matrix $\bTheta$, $\beta_{v,p}$ is the $v$th element of the $V_p$ dimensional vector of coefficients $\bbeta_p$, $\psi_{p,p',h}^a$ denotes the effect of $x_{i,h}$ on the edge connecting $p$th and $p'$th nodes of the network matrix and $\psi_{p,h}^g$ determines the effect of $x_{i,h}$ on the GM image at a voxel within $\mathcal{R}_p$. Considering the symmetry and zero diagonal constraints in the network object $\bA_i$, we set $\theta_{p,p'}=\theta_{p',p}$ and $\theta_{p,p}=0$, for all $1\leq p<p'\leq P$. Additionally, we assume a bilinear structure for network effects for covariates, given by $\psi_{p,p',h}^a=\psi_{p,h}^a\psi_{p',h}^a$, to reduce the parameter space while allowing a sufficiently flexible structure to address potential confounding bias in our primary predictor of interest. 
When both sets of responses follow a normal linear model with an identity link function, (\ref{joint_model}) becomes
\begin{align}\label{joint_model_2}
a_{i,(p,p')} = \sum_{h=1}^{H} \psi_{p,h}^a \psi_{p',h}^a x_{i,h}+\theta_{p,p'}\: y_i+e_{i}^{(p,p')},\:\:
g_{i,v,p} = \sum_{h=1}^{H} \psi_{p,h}^g x_{i,h} +\beta_{v,p}\: y_i+w_{i}^{(v,p)},
\end{align}
for $v=1,\ldots,V_p;\:\:p=1,\ldots,P,$ where $e_{i}^{(p,p')}$ and $w_{i}^{(v,p)}$ represent the errors in the two regression models. 
This article focuses on integrated learning of the mean structure for two sets of models and assumes $e_{i}^{(p,p')}\stackrel{ind.}{\sim} N(0,\tau_{\theta}^2)$ (for $1\leq p<p'\leq P$) and $w_{i}^{(v,p)}\stackrel{ind.}{\sim} N(0,\tau_{\beta}^2)$ (for $v=1,\ldots,V_p$) for simplicity, following the literature on the multivariate linear response regression model \citep{goh2017bayesian}. Consistent with the symmetry and zero diagonal entry in $\bA_i$, we assume $e_{i}^{(p,p')}=e_{i}^{(p',p)}$ and $e_{i}^{(p,p)}=0$. Therefore, stacking over elements of the network matrix and elements of the GM voxels over each region, (\ref{joint_model_2}) can be written as 
\begin{align}\label{joint_model_3}
\bA_i = \sum_{h=1}^{H} \bPsi_h^a x_{i,h} + \bTheta\: y_i + \bE_i,\:\:
\bg_{i,p} = \sum_{h=1}^{H} {\boldsymbol 1}_{V_p}\psi_{p,h}^g x_{i,h} +\bbeta_{p}\: y_i + \bw_i^{(p)},\:\: p=1,\ldots,P,
\end{align}
where $[\bPsi_h^a]_{i,j} = [\bpsi_{h}^a (\bpsi_{h}^{a})']_{i,j}$ for $i \neq j$, $[\bPsi_h^a]_{i,i} = 0$,  $\bpsi_{h}^a =(\psi_{1,h}^a,...,\psi_{P,h}^a)'$ and ${\boldsymbol 1}_{V_p}$ represents a $V_p$-dimensional vector of ones. The error 
$\bE_i\in\mathbb{R}^{P\times P}$ is the symmetric matrix with zero diagonal entries corresponding to the network object and $\bw_i^{(p)}$ is the $V_p$-dimensional error vector corresponding to the GM image at the $p$th ROI. The key to integrated learning of the multi-modal data lies in the development of an integrated prior structure on $\bTheta$ and $\bbeta_p$'s, as described in the next section.

\subsection{Prior Distribution on Multi-Modal Coefficients}\label{sec:prior_coef}
Our integrated prior construction on coefficients $\bbeta_p$'s and $\{\theta_{p,p'}:p<p'\}$ for multi-modal predictors is fundamental to exploiting topology of the image objects and cross-information among them by forming principled linkages among images at the ROI level. The prior construction is aimed at: (a) identification of influential ROIs with uncertainty; (b) shrinkage of unimportant voxel coefficients to zero within an influential ROI; and (c) guaranteeing efficient computation of the posterior for the proposed prior. We cast the problem of identifying influential ROIs from the images as a high-dimensional variable selection problem and formulate prior distributions on multi-modal object coefficients building upon the existing literature on prior constructions for high-dimensional regression coefficients.


The direct application of variable selection prior \citep{george1993variable,carvalho2010horseshoe} to multi-modal coefficients is unappealing for multiple reasons. First, an ordinary variable selection prior on coefficients $\bTheta$ and $\bbeta_p$'s identifies cells in $\bA_i$ and $\bg_{i,p}$ (network edges and GM voxels) significantly related to the predictor, rather than identifying influential ROIs. Second, we seek to impose two additional restrictions on the prior construction of $\bTheta$ and $\bbeta_p$ motivated by the neuro-scientific application. First, if at least one of the $p$th and $p'$th ROIs is not related to the speech rate, the edge coefficient $\theta_{p,p'}$ corresponding to the edge between $p$th and $p'$th nodes is unimportant. Second, if the $p$th ROI is deemed uninfluential then all voxels within the $p$th ROI are unrelated to the speech rate, i.e., $\bbeta_p=\bzero$. These restrictions are relevant due to the hierarchical arrangement of voxels and ROIs and are jointly referred to as the \emph{hierarchical constraint}. Finally, we expect the matrix of coefficients $\bTheta$ (which itself can be regarded as describing a weighted network) to exhibit \emph{transitivity effects}, that is, we expect that if the interactions between regions $p$ and $p'$ and between regions $p'$ and $p''$ are both related to the speech rate, the interaction between regions $p$ and $p''$ will likely be associated with the speech rate. The property of transitivity reflects the assumption of small-world network structure which is a popular model for neural organization that achieves the simultaneous demands of network integration and segregation \citep{bassett} and has been considered in prior studies of PPA \citep{nigro, tao}. An ordinary variable selection prior on multi-modal coefficients does not necessarily conform to all these requirements.

Let $\xi_1,...,\xi_P$ denote the binary inclusion indicators corresponding to the $P$ ROIs taking values in $\{0,1\}$, with $\xi_p=0$ determining no effect of the $p$th ROI on the response from all covariates. In our construction, 
the network edge coefficient $\theta_{p,p'}$ is endowed with a variable selection prior given by
\begin{align}\label{spike-and-slab}
\theta_{p,p'}|\lambda_{p,p'},\tau^2_\theta,\sigma_{\theta},\xi_p,\xi_{p'}\stackrel{ind.}{\sim} \xi_p\xi_{p'}N(0,\tau^2_{\theta}\sigma^2_{\theta}\lambda_{p,p'}^2)+(1-\xi_p\xi_{p'})\delta_0,\:\:p<p',
\end{align}
where $\delta_0$ corresponds to the Dirac-delta function, $\lambda_{p,p'}$ is a local shrinking parameter corresponding to the $(p,p')$th edge and $\sigma_{\theta}$ is the global shrinking parameter for the network coefficient. The prior closely mimics the spike-and-slab variable selection structure with an important difference. While an ordinary spike-and-slab prior introduces a binary inclusion indicator corresponding to each variable, (\ref{spike-and-slab}) enforces $\theta_{p,p'}=0$ when either $\xi_p=0$ or $\xi_{p'}=0$. Such a formulation is sensible from a network perspective as it implies that the edge connecting two network nodes is unrelated to the predictor when at least one of the network nodes is not influential, i.e., it satisfies \emph{hierarchical constraint}. Additionally, the formulation naturally incorporates transitivity effects in the network coefficient $\bTheta$. Further, the use of a single binary inclusion indicator to simultaneously determine the influence of ROIs with respect to both structural and network information is informed by prior evidence that structural atrophy and disruptions of functional connectivity can be co-localized in PPA disease progression \citep{Battistella2019} as well as other progressive NDs \citep{Johnson2000} (see Section 2 for details).  We further assign half-Cauchy distributions on $\sigma_{\theta}\sim C^{+}(0,1)$ and $\lambda_{p,p'}\stackrel{ind.}{\sim} C^{+}(0,1)$ to complete prior specification on network coefficient. Integrating out
$\sigma_{\theta}$ and $\lambda_{p,p'}$ in (\ref{spike-and-slab}), $\theta_{p,p'}|\tau^2_\theta,\xi_p=1,\xi_{p'}=1$ follows the popular horseshoe prior \citep{carvalho2010horseshoe} which offers a flexible prior structure for precise estimation of nonzero network edge coefficients a posteriori.

The GM coefficient $\beta_{v,p}$ in the $p$th ROI is modeled using $\beta_{v,p}=\xi_p\gamma_{v,p}$, to ensure all voxel coefficients in the $p$th ROI become unrelated to the predictor if the $p$th ROI is uninfluential (i.e., $\xi_p=0$), thus satisfying the \emph{hierarchical constraint}. To estimate voxel level effects in the $p$th ROI on the predictor, each $\gamma_{v,p}$ is assigned a horseshoe shrinkage prior which takes the following scale-mixture representation,
\begin{align}\label{GM_prior}
\gamma_{v,p}|\phi_{v,p},\eta_p^2,\tau^2_\beta \sim N(0, \tau^2_\beta \eta_p^2 \phi_{v,p}^2),\:\:\phi_{v,p}\stackrel{i.i.d.}{\sim} C^{+}(0,1),\:\:\eta_p\stackrel{i.i.d.}{\sim} C^{+}(0,1),
\end{align}
for $v=1,...,V_p;\:p=1,...,P$. The prior structure (\ref{GM_prior}) induces approximate sparsity in voxel-level GM coefficients $\gamma_{v,p}$ by shrinking the components which are less influential toward zero while retaining the true signals \citep{polson2010shrink}.  Finally, the binary inclusion indicators are assigned Bernoulli prior distribution $\xi_p\stackrel{i.i.d.}{\sim} Ber(\nu)$ with $\nu\sim Beta(a_{\nu},b_{\nu})$ to account for multiplicity correction. Notably, an estimate of the posterior probability of the event $\{\xi_p=1\}$ shows the uncertainty in identifying the $p$th ROI to be influential. Thus, $P(\xi_p=1|\mbox{Data})$ close to $1$ or $0$ signifies strong evidence in favor of identifying the $p$th ROI to be active or inactive, respectively. The prior specification on multi-modal coefficients is completed by assigning an inverse Gamma $IG(a_{\tau_\theta},b_{\tau_\theta})$ and $IG(a_{\tau_\beta},b_{\tau_\beta})$ prior on the error variances $\tau^2_\theta$ and $\tau^2_\beta$ respectively. We assign a flat prior on $\psi_{p,h}^a,\psi_{p,h}^g$
for all $p=1,...,P$ and $h=1,...,H$.

Notably, the integrated prior distribution ensures significant feedback from voxel-level inference to ROI-level inference. An ROI is determined to be influential by considering the relevance of all the voxels within that ROI, along with the relevant edges of the network object. This means that multiple weak voxel-level signals within an ROI can collectively identify it as influential, or a strong signal at a single voxel can make the ROI influential. Our method achieves a principled Bayesian balance, weighing the strength of all signals contributing to that ROI. We study these aspects empirically in Section~\ref{sec:sim_study}.
Additionally, detailed inference on $\gamma_{v,p}$ enables voxel-level association between the image and the speech rate measure from our method for influential ROIs. Given that the $\gamma_{v,p}$ values are assigned a continuous shrinkage prior, we need to apply post-processing on the MCMC samples of $\gamma_{v,p}$ to determine influential voxels following the approach described in the literature \citep{li2017variable}. While we present inference on $\beta_{v,p}$, we do not emphasize voxel identification in the simulation study following the post-processing steps with $\gamma_{v,p}$ due to space constraint, since our main focus is identifying influential ROIs.



\section{Posterior Computation}\label{posterior_computation}
Full conditional distributions for all
the parameters are available and mostly correspond to standard
families (available in the supplementary file). Thus, posterior computation can proceed through a Markov chain Monte Carlo (MCMC) algorithm. 
The MCMC sampler is run for $5000$ iterations, with the first $1000$ discarded as burn-in. 
All simulation scenarios show an average effective sample size of over 3143, indicating fairly uncorrelated post-burn-in MCMC samples.
Different
replications of the model are implemented under a parallel
architecture by making use of the packages \texttt{doparallel} and \texttt{foreach} within R.


$L$ (suitably thinned) post-convergence MCMC samples $\xi_p^{(1)},...,\xi_P^{(L)}$ of the binary indicator $\xi_p$ are used to empirically assess if the $p$th ROI is significantly associated with the response. In particular, the $p$th ROI $\mathcal{R}_p$ is related to the response if  $\sum_{l=1}^L\xi_p^{(l)}/L>t$, for $0<t<1$. The ensuing simulation section computes the $F_1$ score for $t=0.5$ to decide which ROIs are influential in predicting the response.

One important point to note is that $\sum_{l=1}^L\xi_p^{(l)}/L$ empirically approximates $P(\xi_p=1|\mbox{Data})$, and offers the uncertainty of identifying the $p$th ROI to be influential. Offering uncertainty in region selection is a salient feature of the proposed model. To determine whether a region is influential, a threshold is required. We have adopted the median probability rule \citep{barbieri2004optimal} of $t=0.5$, which is widely adopted in the literature. However, we acknowledge that this cut-off may not always be optimal when variables are highly correlated \citep{barbieri2021median}, a consideration that lies beyond the scope of this article.
Since there is no theoretically or experimentally validated cut-off specific to our model, this choice was the most natural. Nonetheless, alternative thresholds may be selected based on the scientific context of a given study.

\begin{figure}[hbt!]
    \caption{Runtime in seconds for varying number of regions $P$ and number of voxels per region $V$ for fixed $n=25$.}
    \label{fig_runtime}
    \begin{center}
        \includegraphics[trim = 0in 0in 0in 0in, clip, width=0.6\linewidth]{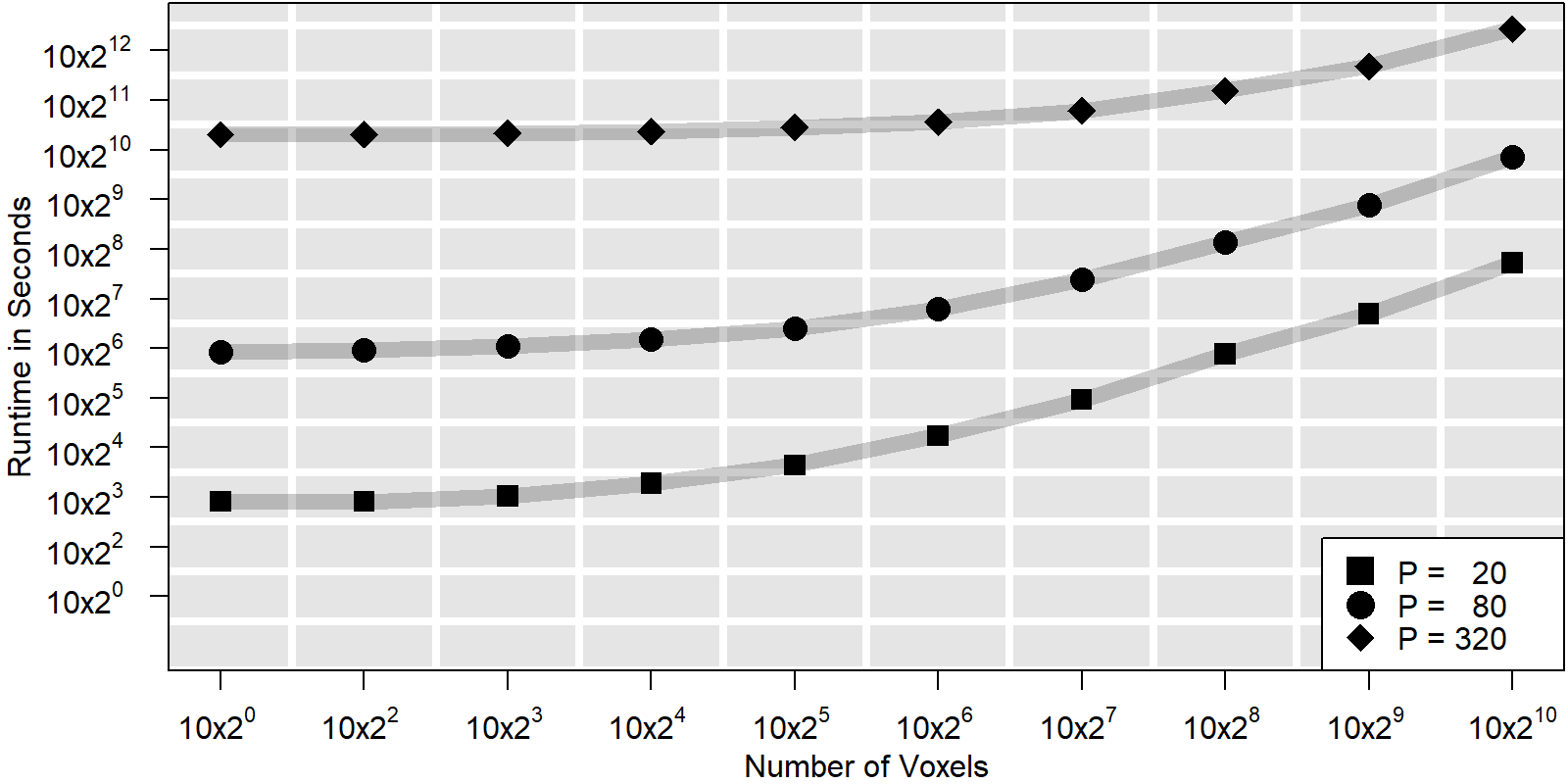}\\
    \end{center}
\end{figure}

Our computational analysis indicates that the algorithm has a complexity of $\mathcal{O}(n^3 P (P + V))$, where $V=V_p$ denotes the number of voxels per region, assuming each region contains an equal number of voxels. With a limited sample size $n$, $P$ and $V$ are the primary drivers of computational complexity. Figure \ref{fig_runtime} illustrates the computation time as the number of regions $P$ and voxels per region $V$ vary, with the sample size fixed at $n=25$. The figure demonstrates that the algorithm exhibits linear complexity with respect to $V$ and provides insights into how computation time increases with a growing number of regions $P$ while $V$ remains constant.


\section{Simulation Studies}\label{sec:sim_study}

This section compares our proposed approach, referred to as Bayesian multi-object response regression (BMRR), to that of a few representative frequentist and Bayesian competitors in terms of identifying influential regions and drawing inference on regression coefficients.

\subsection{Data Generation}

In all our simulations, we assume 3 biological or demographic covariates and a primary covariate of interest. The covariate of interest $y$ is simulated from $N(0,1)$, while we simulate $x_{1}$ form a Bernoulli$(0.5)$, and simulate $x_{2}$ and $x_3$
from $N(0,1)$. The coefficients are then simulated as $\bpsi_{p,.t}^a \stackrel{i.i.d.}{\sim} N(\bzero_H, I_H)$ and $\bpsi_{p,.t}^g \stackrel{i.i.d.}{\sim} N(\bzero_H, I_H)$ for 
$p=1,\ldots,P$. We then generate responses $\bA_i$ and $\bg_{i,p}$ from model (\ref{joint_model_3}) with the true network coefficient $\bTheta_t$ and true structural coefficient $\bbeta_{p,t}$ for the $p$th ROI. The subscript $t$ indicates the true data-generating parameters. In all simulations, we set the sample size to $n=25$ to assess the performance of our approach in a limited sample setting which matches the scenario in our application. We also assume an equal number of voxels per ROI, i.e., $V_1=\cdots=V_P=V$. 

\noindent\underline{\textbf{(A1) Non-spatial simulations: Simulating true coefficients $\bTheta_t$ and $\bbeta_{p,t}$.}} To simulate the true coefficients $\bTheta_t$ and $\bbeta_{p,t}$ without any spatial association between elements of $\bbeta_{p,t}$, we first generate true activation indicators $\{\xi_{p,t}^\theta\}_{p=1}^P$ for each region corresponding to the network object and $\{\xi_{p,t}^\beta\}_{p=1}^P$ for each region corresponding to the structural object. The two sets of indicators can impose varying degrees of co-localization between the modalities, as described below.
\begin{enumerate}
        \item Generate $\kappa_{p,t} \sim Bernoulli(\nu_t c_t)$, where $0\leq \nu_t\leq 1$ and $0\leq c_t\leq 1$.
        \item Generate $\{\xi_{p,t}^\theta\}_{p=1}^P$ as follows:
        $$\xi_{p,t}^\theta = \begin{cases} 1               & \text{if} \quad \kappa_{p,t} = 1 \\
                                   \hat{\xi}_{p,t}^\theta  & \text{if} \quad \kappa_{p,t} = 0
                            \end{cases}, \:\:\hat{\xi}_{p,t}^\theta \sim Bernoulli(\nu_t \frac{1 - c_t}{1 - \nu_t c_t})$$
        where $\hat{\xi}_{p,t}^\theta$ indicates activation for region $p$ of the network object independently of the structural object.
        \item Generate $\{\xi_{p,t}^\beta\}_{p=1}^P$ as follows:
        $$\xi_{p,t}^\beta = \begin{cases} 1              & \text{if} \quad \kappa_{p,t} = 1 \\
                                   \hat{\xi}_{p,t}^\beta & \text{if} \quad \kappa_{p,t} = 0
                        \end{cases},\:\:\hat{\xi}_{p,t}^\beta \sim Bernoulli(\nu_t \frac{1 - c_t}{1 - \nu_t c_t})$$
        where $\hat{\xi}_{p,t}^\beta$ indicates activation for region $p$ of the structural object independently of the network object.
        \item $p$th region is defined to be influential if either $\xi_{p,t}^\theta=1$ or $\xi_{p,t}^\beta=1$. Thus, if   $\xi_{p,t}$ denotes the true activation of region $\mathcal{R}_p$, by definition
        $\xi_{p,t}=1$  if $\xi^\theta_{p,t} = 1$ or $\xi^\beta_{p,t} = 1$.
\end{enumerate}

Note that $c_t = 1$ implies $\xi_{p,t} = \xi^\theta_{p,t} = \xi^\beta_{p,t}$. We refer to this case as \emph{perfect co-localization}. In general, the indicators $\xi^\theta_{p,t},  \xi^\beta_{p,t}$ will be different, with more coincidences as $c_t$ is close to $1$, and less coincidences as $c_t$ is close to $0$. Importantly, even if $c_t=0$, there is a nonzero probability of $\xi_{p,t} = \xi^\theta_{p,t} = \xi^\beta_{p,t}$ for a region $\mathcal{R}_p$. Since $\mathbb{E}[\xi_p^\theta]=\mathbb{E}[\xi_p^\beta]=\nu_t$, $(1-\nu_t)$ controls the probability of a region not being ``influential". It is known as the \emph{node sparsity parameter}.


For each sparsity level and co-localization degree, the coefficient corresponding to the edge connecting the $p$-th and $p'$-th region is drawn from the mixture distribution
\begin{align}\label{true_network_coeff}
\theta_{(p,p'),t}|\xi^\theta_{p,t},\xi^\theta_{p',t}\sim \xi_{p,t}^{\theta}\xi_{p',t}^{\theta}N(\mu_{\theta,(p,p')},\sigma_{\theta}^2)+ 
(1-\xi_{p,t}^{\theta}\xi_{p',t}^{\theta})\delta_0,\:\:\:\theta_{(p,p'),t}=\theta_{(p',p),t}.
\end{align}
(\ref{true_network_coeff}) ensures that any edge connecting to the $p$-th region in the network response is unrelated to the predictor if the $p$-th region is un-influential, i.e., $\xi^\theta_{p,t}=0\Rightarrow \theta_{(p,p'),t}=0$ for all $p'\in\{1,..,P\}$. Similarly, for each un-influential region $\mathcal{R}_p$, i.e., when $\xi^\beta_{p,t} = 0$, the $V$ dimensional GM coefficient $\bbeta_{p,t}$ is set at $\bzero$. When $\xi^\beta_{p,t}=1$, i.e., the $p$-th region is influential, we randomly choose $\upsilon_t$ proportion of cell coefficients in the $p$-th region to be nonzero and the rest are set at zero. Thus $\upsilon_t$ is the density of activated voxels within a region. The nonzero coefficients within $\bbeta_{p,t}$ are simulated from $N(\mu_{\beta,(v,p)},\sigma_{\beta}^2)$, with $\mu_{\theta,(p,p')}$ and $\mu_{\beta,(v,p)}$ are drawn from $Unif(0.25, 1)$. We fix $\sigma_{\theta}^2=1$, $\sigma_{\beta}^2=1$. We control the signal-to-noise ratio by changing the error variances $\tau_{t,\theta}^2$ and $\tau_{t,\beta}^2$.

\noindent\underline{\textbf{(A2) Spatial simulations: Simulating true coefficients $\bTheta_t$ and $\bbeta_{p,t}$.}}
We also consider simulation scenarios where structural coefficients are spatially indexed, i.e., $\beta_{v,p} = \beta(\bs_{v,p})$, with $\bs_{v,p}$ representing the spatial location of the $v$th voxel in the $\mathcal{R}_p$. 
This simulation setting assumes \emph{perfect co-localization}, i.e., $\xi_{p,t}=
\xi_{p,t}^\theta=\xi_{p,t}^\beta$. The network edge coefficient $\theta_{(p,p'),t}$ is simulated following (\ref{true_network_coeff}). When $\xi_{p,t}=1$, we randomly choose $\upsilon_t$ proportion of voxel coefficients in $\mathcal{R}_p$ to be nonzero, and the rest are set at zero. the nonzero coefficient $\beta(\bs_{v,p})$ is simulated from a Gaussian process with mean 1 and a covariance function given by the Matérn Kernel:
$cor(\beta(\bs_{v,p}),\beta(\bs_{v',p}))=K(d|\rho,\eta) = \frac{2^{1-\eta}}{\Gamma(\eta)} \left((2\eta)^{1/2}\frac{d}{\rho}\right)^\eta K_\eta\left((2\eta)^{1/2}\frac{d}{\rho}\right),$
where $d = ||\bs_{v,p} - \bs_{v',p}||$ and $K_{\eta}$ is the modified Bessel function of order $\eta$. We set $\eta=2$ to ensure a fairly smooth spatial surface and $\rho=1$ for a reasonable range of spatial dependence.

We simulate data under different node sparsity parameters $(1-\nu_t)$, different signal-to noise ratio, different degree of co-localization $c_t$ and different density of activated voxels $\upsilon_t$ within an ROI. 
Finally, a different number of ROIs and different numbers of voxels within an ROI are considered. Specifically, we consider two cases, (a) $P=20$ and $V=10$, and  (b) $P = 100$ and $V = 50$. Cases (a) and (b) are referred to as the ``small dimensional example" and ``high dimensional example."
 All simulation scenarios are summarized in Table~\ref{tab_col_sim_esc}. 

 \begin{table}
\caption{Simulation scenarios by varying degree of co-localization ($c_t$), node sparsity parameter ($1-\nu_t$), signal-to-noise ratio, number of ROIs and number of voxels per ROI, and the density of activated voxels per region ($\upsilon_t$).}
\label{tab_col_sim_esc}
\centering
\tiny
\begin{tabular}[t]{llcccccc}
\toprule
\multicolumn{2}{c}{} & \multicolumn{3}{c}{node sparsity $(1-\nu_t)=$ 0.85} & \multicolumn{3}{c}{node sparsity $(1-\nu_t)=$ 0.70} \\
\cmidrule(l{3pt}r{3pt}){3-5}
\cmidrule(l{3pt}r{3pt}){6-8}
\multicolumn{1}{c}{} & \multicolumn{1}{c}{$SNR$}   & 
                        \multicolumn{1}{c}{$c_t=1$}   &            
                       \multicolumn{1}{c}{$c_t=0.5$} &            
                       \multicolumn{1}{c}{$c_t=0$}   &            
                       \multicolumn{1}{c}{$c_t=1$}   &            
                       \multicolumn{1}{c}{$c_t=0.5$} &            
                       \multicolumn{1}{c}{$c_t=0$}   \\
\midrule
Low   & $4$    & $\upsilon_t=(0.05,..,1)$ & $\upsilon_t=(0.05,..,1)$ & $\upsilon_t=(0.05,..,1)$ & $\upsilon_t=(0.05,..,1)$ & $\upsilon_t=(0.05,..,1)$ & $\upsilon_t=(0.05,..,1)$ \\
Dim.  & $2$    & $\upsilon_t=(0.05,..,1)$ & $\upsilon_t=(0.05,..,1)$ & $\upsilon_t=(0.05,..,1)$ & $\upsilon_t=(0.05,..,1)$ & $\upsilon_t=(0.05,..,1)$ & $\upsilon_t=(0.05,..,1)$ \\ 
      & $1$    & $\upsilon_t=(0.05,..,1)$ & $\upsilon_t=(0.05,..,1)$ & $\upsilon_t=(0.05,..,1)$ & $\upsilon_t=(0.05,..,1)$ & $\upsilon_t=(0.05,..,1)$ & $\upsilon_t=(0.05,..,1)$ \\
      & $0.5$  & $\upsilon_t=(0.05,..,1)$ & $\upsilon_t=(0.05,..,1)$ & $\upsilon_t=(0.05,..,1)$ & $\upsilon_t=(0.05,..,1)$ & $\upsilon_t=(0.05,..,1)$ & $\upsilon_t=(0.05,..,1)$ \\
      & $0.25$ & $\upsilon_t=(0.05,..,1)$ & $\upsilon_t=(0.05,..,1)$ & $\upsilon_t=(0.05,..,1)$ & $\upsilon_t=(0.05,..,1)$ & $\upsilon_t=(0.05,..,1)$ & $\upsilon_t=(0.05,..,1)$ \\
\midrule
High   & $4$    & $\upsilon_t=(0.05,..,1)$ & $\upsilon_t=(0.05,..,1)$ & $\upsilon_t=(0.05,..,1)$ & $\upsilon_t=(0.05,..,1)$ & $\upsilon_t=(0.05,..,1)$ & $\upsilon_t=(0.05,..,1)$ \\
Dim.  & $2$    & $\upsilon_t=(0.05,..,1)$ & $\upsilon_t=(0.05,..,1)$ & $\upsilon_t=(0.05,..,1)$ & $\upsilon_t=(0.05,..,1)$ & $\upsilon_t=(0.05,..,1)$ & $\upsilon_t=(0.05,..,1)$ \\ 
      & $1$    & $\upsilon_t=(0.05,..,1)$ & $\upsilon_t=(0.05,..,1)$ & $\upsilon_t=(0.05,..,1)$ & $\upsilon_t=(0.05,..,1)$ & $\upsilon_t=(0.05,..,1)$ & $\upsilon_t=(0.05,..,1)$ \\
      & $0.5$  & $\upsilon_t=(0.05,..,1)$ & $\upsilon_t=(0.05,..,1)$ & $\upsilon_t=(0.05,..,1)$ & $\upsilon_t=(0.05,..,1)$ & $\upsilon_t=(0.05,..,1)$ & $\upsilon_t=(0.05,..,1)$ \\
      & $0.25$ & $\upsilon_t=(0.05,..,1)$ & $\upsilon_t=(0.05,..,1)$ & $\upsilon_t=(0.05,..,1)$ & $\upsilon_t=(0.05,..,1)$ & $\upsilon_t=(0.05,..,1)$ & $\upsilon_t=(0.05,..,1)$ \\
\bottomrule
\end{tabular}
\end{table}

\subsection{Competitors and Metrics of Comparison}\label{sec:competitors}
The simulated data will be used to assess the performance of: (A) identifying influential regions; (B) estimating the true network coefficient $\bTheta_t$ and structural coefficients $\bbeta_{1,t},...,\bbeta_{P,t}$; and (C) quantifying uncertainty in inference. We construct a series of competitors to assess (A)-(C) as below.

\subsubsection{Frequentist competitors}
As frequentist competitors, we implement popularly used mass univariate analysis (MUA). In this approach each network edge and each cell of the structural image is regressed on the predictor of interest, along with biological and demographic covariates via multiple linear regression with a bi-linear structure on the coefficients to obtain p-values corresponding to the point estimates of $\theta_{p,p'}$ and $\beta_{v,p}$, denoted by p-value($\hat{\theta}_{p,p'}$) and p-value($\hat{\beta}_{v,p}$), respectively. 
These p-values will be compared to a threshold to declare if a region is influential, accounting for the multiple comparison issues by controlling the False Discovery Rate (FDR) \citep{genovese_thresholding_2002}. Specifically, 
MUA orders p-value($\hat{\beta}_{v,p}$) and p-value($\hat{\theta}_{p,p'}$) for all $v=1,...,V_p$, $1\leq p<p'\leq P$ in ascending order. Let $\text{p-value}_{(1)},...,$ $\text{p-value}_{\left(\frac{P(P-1)}{2} + \sum_{p=1}^P V_p \right)}$ be the ordered p-values and let $i^*$ be the largest index such that $\text{p-value}_{(i^*)} \leq \frac{i^*\alpha_0}{\frac{P(P-1)}{2} + \sum_{p=1}^P V_p}$. A region $p$ is identified as influential if $\text{p-value}(\hat{\theta}_{p,p'}) \leq \text{p-value}_{(i^*)}$ for at least one $p' \neq p$, or $\text{p-value}(\hat{\beta}_{v,p}) \leq \text{p-value}_{(i^*)}$ for at least one $v\in\{1,...,V_p\}$. We set $\alpha_0=0.05$ throughout this discussion. 
We also implement tensor response regression (TRR) \citep{li2017parsimonious} where both objects are viewed as 2D tensors in the simulation. TRR is implemented using the R package \texttt{TRES}. 
A region is considered influential by TRR if either an edge connected to that region or a voxel in that region is influential.


\subsubsection{Bayesian competitors}
We construct a few Bayesian competitors to compare single modal analysis and joint analysis of modalities without integrating information between objects, with the integrative analysis performed by BMRR. These competitors include:
(a) fitting only the Bayesian network on scalar regression (i.e., the first equation in (\ref{joint_model_3})), referred to as the \emph{network response regression} (NRR); (b) fitting only the Bayesian structural images on scalar regression (i.e., the second equation in (\ref{joint_model_3}), referred to as the \emph{structural response regression} (SRR); and (c) fitting both of them jointly without integrating them through shared parameters, referred to as the \emph{joint regression} (JR). For NRR, a region is identified as influential if at least one edge connecting to that region is influential. In SRR, a region is deemed influential if it contains at least one influential voxel. For JR, a region is identified as influential if any voxel in that region or any edge connected to that region is influential.

We implement two additional Bayesian competitors, which do not acknowledge the network topology or the connection between the topology of two sets of objects through the hierarchical constraint. The first Bayesian competitor, referred to as the Spike \& Slab, applies an ordinary spike \& slab prior \citep{george1993variable} on each $\theta_{p,p'}$ and $\beta_{v,p}$. 
We also implement a Horseshoe shrinkage prior \citep{carvalho2010horseshoe} on each $\theta_{p,p'}$ and $\beta_{v,p}$. As in the frequentist competitors, we controlled for False Discovery Rate (FDR). Unlike BMRR, which naturally selects regions, these methods require careful tuning for FDR-based region selection. For the Spike \& Slab method, FDR is applied using the indicator variables. However, it is not straightforward for the Horseshoe prior and it requires a rejection region. Since we could not empirically determine this region, we used nested rejection regions (intervals) for individual coefficients, selecting the smallest interval that met the FDR threshold. We control FDR for joint objects similar to the frequentist methods, yielding the best performance.

It is important to highlight that while the Horseshoe prior provides strong theoretical guarantees for estimation and uncertainty quantification of regression coefficients \citep{van2017uncertainty}, its theoretical foundation for identifying important predictors remains limited. In particular, the variable selection framework using the Horseshoe prior in our joint regression setting lacks theoretical guarantees for influential region identification with uncertainty quantification. This is a key advantage of BMRR, which not only facilitates region selection but also provides precise uncertainty quantification in this process.

\subsubsection{Metrics of comparison}\label{sec:metrics}
We will present the $F_1$ score computed from the True Positive Rate (TPR) and the True Negative Rate (TNR) for correctly identifying important regions.
The point estimation of every competitor is assessed using mean squared errors (MSE) of estimating the network coefficient $\bTheta_t$ and the structural coefficients $\bbeta_{1,t},...,\bbeta_{P,t}$. Since both the fitted $\bTheta$ and $\bTheta_t$ are symmetric with zero diagonals, the 
MSE for both sets of coefficients jointly is given by $[\sum_{p<p'}(\theta_{p,p',t}-\widehat{\theta}_{p,p'})^2+\sum_{p=1}^P||\bbeta_{p,t}-\widehat{\bbeta}_p||^2]/(P(P-1)/2+VP)$. The point estimates are taken to be the posterior median for the Bayesian competitors.
For Bayesian competitors, we evaluate the length and coverage of 95\% credible intervals averaged across coefficients in $\bTheta$ and $\bbeta_{p}$'s. All results presented over $500$ simulated datasets.

\subsection{Results}\label{sec:results}
\begin{figure}[hbt!]
    \caption{$F_1$ score for BMRR under simulation scenarios outlined in Table~\ref{tab_col_sim_esc}.}
    \label{fig_f1_sco}
    \begin{center}
        \includegraphics[trim = 0in 0in 0in 0in, clip, width=0.7\linewidth]{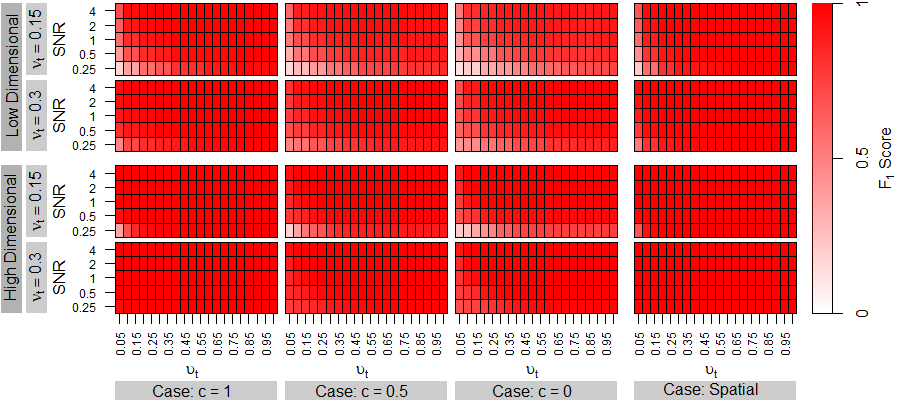}\\
    \end{center}
\end{figure}

Figure \ref{fig_f1_sco} presents the $F_1$ score for BMRR region identification under each of our simulation scenarios in Table~\ref{tab_col_sim_esc}. As the voxel density parameter $\upsilon_t$ approaches 0, indicating high voxel-sparsity, region identification suffers. Similarly, BMRR's performance deteriorates with lower SNR. However, for the same $\upsilon_t$ and SNR, BMRR performs better in high-dimensional cases than in low-dimensional cases. Additionally, for the same levels of $\upsilon_t$, SNR, and $(V,P)$, BMRR performs better with a higher degree of co-localization, since it induces co-localization in its modeling framework. Finally, smoothly varying spatial correlation over the voxels of the structural images does not significantly impact BMRR's performance in terms of region identification.

\begin{figure}[hbt!]
    \caption{$F_1$ score for different simulation scenarios for BMRR integrating the objects, using both objects without integration (JR), only the structural object (SRR), and only the network object (NRR).}
    \label{fig_obj_plo}
    \begin{center}
        \includegraphics[trim = 0in 0in 0in 0in, clip, width=0.75\linewidth]{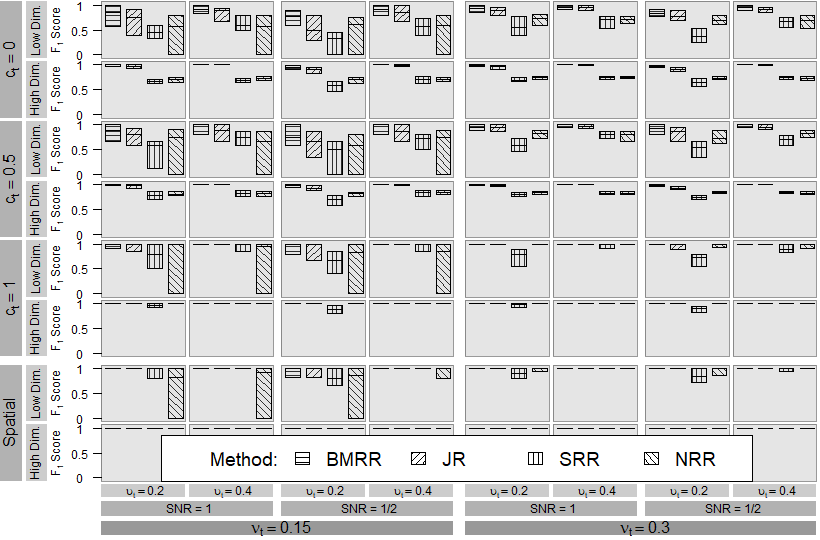} \\
    \end{center}
\end{figure}

Figure~\ref{fig_obj_plo} shows boxplots of $F_1$ scores for BMRR, NRR, SRR, and JR. Due to space constraints, we present results for moderate and low SNR values of $1$ and $0.5$, and voxel density parameter $\upsilon_t = 0.2$ and $0.4$. The integrated version, BMRR, consistently outperforms using either or both objects independently. While using both objects in JR improves performance compared to using only one image in NRR or SRR, the improvement is not as significant as when using the integrated BMRR approach. Both BMRR and JR show marginal performance improvements with increasing SNR, increasing co-localization degree, and in high-dimensional settings. In contrast, the improvement is significant for NRR and SRR in the high-dimensional setting. All competitors show better performance with increasing voxel density as well as with decreasing node sparsity within a limit. Additionally, the $F_1$ scores do not show any significant perturbation with spatial association between voxels of the structural image.


Our focus now lies on the performance of BMRR compared to other competitors excluding the Bayesian structured regressions. Due to space restrictions, we only consider representative simulation cases where SNR is $1$, node sparsity $(1-\nu_t)=0.7$, and 
voxel density $\upsilon_t=0.4$. Results for other simulation cases lead to similar conclusions.

\begin{figure}[hbt!]
    \caption{$F_1$ score, MSE, and $95\%$ interval coverage and length for BMRR and frequentist and Bayesian competitors. The interquartile range and median across simulations are presented.}
    \label{fig_com_sta}
    \begin{center}
        \includegraphics[trim = 0in 0in 0in 0in, clip, width=0.85\linewidth]{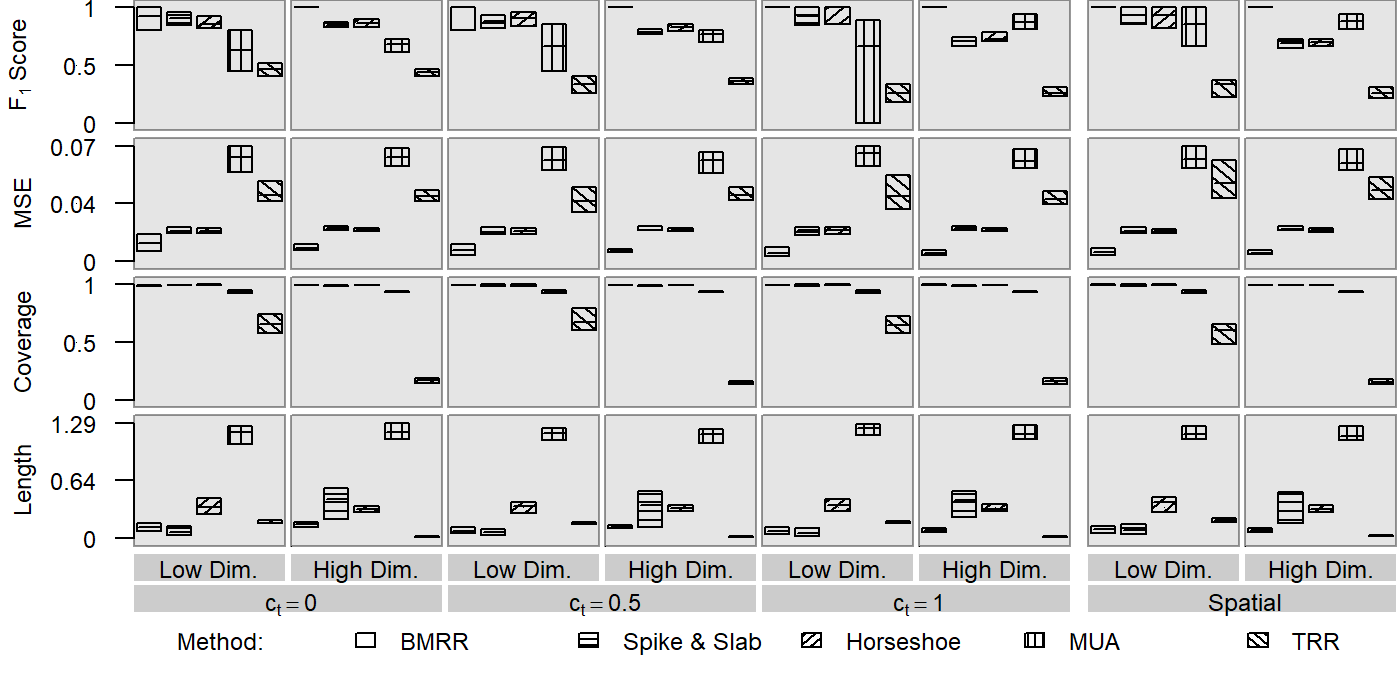} \\
    \end{center}
\end{figure}

According to Figure \ref{fig_com_sta}, BMRR outperforms every competitor in every scenario in terms of $F_1$ score. While Spike \& Slab, Horseshoe, and MUA show average $F_1$ scores close to 1 as the co-localization degree increases, MUA scores exhibit wide variability over replications, as demonstrated by the large interquartile ranges. BMRR also outperforms other competitors in terms of estimating network and structural coefficients, as evidenced by the smallest MSE values under all scenarios. Spike \& Slab and Horseshoe perform similarly to each other, while MUA and TRR show the poorest performance. Notably, the performance of all competitors remains largely unchanged between spatial and non-spatial scenarios with perfect co-localization. For uncertainty quantification, we observe that all methods offer nominal or slight over-coverage, except for TRR, which shows severe under-coverage. Additionally, BMRR provides narrower 95\% credible intervals than its competitors under all scenarios, with the exception of Spike \& Slab, which marginally outperforms BMRR in low-dimensional cases.

Notably, our simulation study explores the impact of violating the co-localization assumption on model performance.  While we observe a decrease in region identification performance (F1-score), the effects on mean squared error (MSE), coverage, and interval length are less pronounced.  Importantly, our method still outperforms other approaches even when co-localization is not strictly met. This demonstrates that while co-localization is a desirable characteristic, it is not a prerequisite for effective performance with our method.  These results support our scientific motivation of achieving accurate simultaneous region identification across both modalities, even in the presence of imperfect co-localization.

\section{Application to the multi-modal PPA data}\label{PPA_analysis}

\subsection{Data structure and methods}

In our motivating study, sMRI GM, resting state fMRI connectivity images, and speech rates (described in Section 2) were collected at the University of California, San Francisco Memory and Aging Center on 24 subjects with nfvPPA (10 males and 14 females) aged between 57 and 81 years (68.9 years old on average). sMRI GM maps were preprocessed as in \cite{canu} and fMRI network images were processed as in \citet{montembeault}. 
We adjust our analysis for biological sex, age, and cognitive function measured via the mini-mental state examination (MMSE) scale as a measure of dementia severity as in \cite{garcia}.

The sMRI GM segmentation probabilities and fMRI Pearson correlations were transformed via inverse normal cumulative distribution function and Fisher Z-transformation, respectively, to produce z-scores that satisfy assumptions of the error distribution. The sMRI GM and fMRI network connectivity images were each mean centered for a fixed voxel and edge across subjects, respectively. The biological sex indicator is encoded as 1 for male and 0 for female. The covariates MMSE and age are mean-centered. Finally, the speech rate predictor variable has been normalized to have a mean of zero and a standard deviation of one. 
We set out to compare the inferential outcomes from BMRR to the best-performing competitors in the simulations, i.e., MUA, Horseshoe, JR, and TRR. Unfortunately, \texttt{TRES} package sends an error message while implementing TRR, which precludes its usage as a competitor.


\subsection{Data analysis results} 

We present the results from our application of the proposed BMRR to the multi-modal PPA imaging data. In total, 46 out of 245 ROIs have a posterior probability $P(\xi_p=1|\mbox{Data})>0.5$ of being associated with speech rate in the multi-modal data. The set of influential ROIs and their posterior probability of being influential (which is a measure of uncertainty) are recorded in Table~\ref{tab_sel_prob_bmrr}. In general, the ROIs lie in areas of the brain previously associated with language, motor speech, and neurodegeneration in PPA, specifically the frontal lobe including the superior frontal gyrus (SFG) \citep{landin}, the middle frontal gyrus (MFG) \citep{landin}, the inferior frontal gyrus (IFG) \citep{garcia}, the inferior parietal lobe (IPL) \citep{bzdok}, the superior parietal lobe (SPL) \citep{alahmadi}, and the basal ganglia (BG) \citep{mont} (see \cite{brainnetome} for Brainnetome atlas ROI naming conventions). The 46 ROIs display a bilateral distribution which is consistent with previous findings that indicate that articulation can be impacted by disruptions in the motor speech network in either the left or right hemisphere of the motor speech network \citep{landin}. 

\begin{figure}[!ht]
 \centering
  \caption{Point estimates (median) of the regression coefficients for gray matter segmentation probability z-scores for selected voxels across horizontal slices using BMRR with speech rate as predictor.}\label{fig_cov_gm_coe_WPM.T}
  \includegraphics[width=0.60\linewidth]{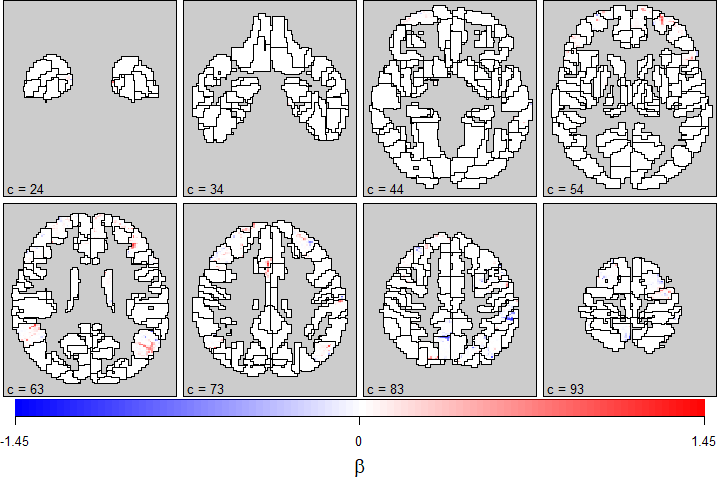}
\end{figure}

Figure \ref{fig_cov_gm_coe_WPM.T} displays horizontal slices (higher c coordinate refers to the top of the brain) of the GM coefficient map organized by ROI where the colored shading indicates the magnitude and direction of association of the speech rate with z-score GM segmentation probability. Due to the hierarchical nature of the BMRR model, the distribution of influential GM voxels (as captured by the coefficients) is determined by the ROI selection and thus reflects the distribution of the selected 46 ROIs discussed above, with concentrations in the SFG, MFG, IFG, IPL, SPL, and BG in a bilateral distribution. The GM coefficient maps provide an additional level of granularity by characterizing the association between the speech rate and GM probability at the voxel level as well as the sparsity of the signal which varies by ROI. Overall, speech rate is positively associated with increases in voxel level GM in the frontal gyrus which affirms that a decrease in GM in these regions is correlated with lower articulation rates \citep{landin, garcia}. However, there is heterogeneity in the voxel level signal both within and across ROIs with a substantial number of voxels displaying a negative association of GM with speech rate particularly in the parietal lobe. The effect sizes range from -1.4 to 1.4 across voxels which indicates that a standard deviation shift in speech rate (25 words/minute) can produce a change in z-score GM segmentation probability from between -1.4 and 1.4. These results reveal a dynamic portrait of how GM and neurodegeneration are associated with motor speech and highlight heterogeneity in the direction of the association as well as varying levels of sparsity in influential voxels across ROIs, insights that are missed when data is pooled at the ROI level. 

\begin{figure}[!ht]
 \centering
  \caption{Point estimates (median) of the regression coefficients for Network Matrix z-scores using BMRR with speech rate as predictor. Each cell displays the ROI level network coefficient maps organized by anatomical brain region where the top and bottom half of each anatomical region indicated by a horizontal dash corresponds to the left hemisphere and right hemispheres, respectively.}\label{fig_cov_nm_coe_WPM.T}
  \includegraphics[width=0.60\linewidth]{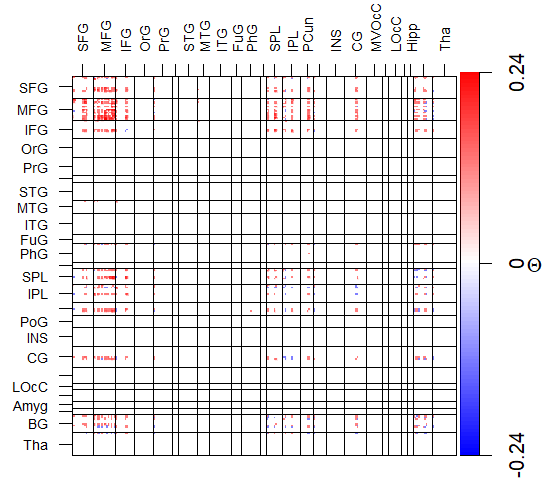}
\end{figure}

Figure \ref{fig_cov_nm_coe_WPM.T} displays the ROI level network coefficient maps organized by anatomical brain region where the top and bottom half of each anatomical region indicated by a horizontal dash corresponds to the left hemisphere and right hemispheres, respectively. The colored shading indicates the magnitude and direction of association of the speech rate with resting state functional connectivity (measured as the Fisher Z-transformed pairwise Pearson correlation). The network edges associated with the speech rate are sparse and there is almost always a positive association between functional connectivity and speech rate. There is a high degree of bilateral functional connectivity associated with speech rate both among and between the SFG, MFG, and IFG brain regions in the frontal lobe. Connections extend outside the frontal lobe and higher speech rates are also associated with higher functional connectivity between the frontal lobe and the parietal lobe which has been indicated in higher-order language processing \citep{coslett}, including the SPL, IPL, and the precuneus (PCun). The effect sizes range from -0.01 to 0.24 across network edges which indicates that a standard deviation shift in speech rate (25 words/minute) can produce a change in Fisher Z-transformed correlation between -0.01 to 0.24. Supplementary Figures 1-3 show the GM coefficient maps for the covariates gender, age, and MMSE, and Figures 4-6 show the same for network coefficient maps and are interpreted in a similar manner to the speech rate coefficient maps, albeit without the hierarchical regularization priors.

We compare the results from the application of the proposed BMRR to the MUA. Figures 7-8 in the supplementary file show that very few GM voxels and network edges are selected, providing us with very few signals for analysis in MUA. 
Table \ref{tab_pplc} shows better model fit by BMRR than JR in terms of the Posterior Predictive Loss Criterion (PPLC), indicating the advantages of integrating information from objects. Given that the coefficients for the speech rate measure are not very large, BMRR has only a modest advantage over Horseshoe in model fitting statistics.
However, Horseshoe proves extremely unreliable in identifying influential ROIs, as it marks all ROIs as influential, rendering the results scientifically uninterpretable. 

In our empirical analysis, we set the hyper-parameter values as
 $a_\nu = b_\nu = 1 = a_{\tau_\theta} = b_{\tau_\theta} = a_{\tau_\beta} = b_{\tau_\beta} = 1$. To assess the sensitivity of hyper-parameters, we vary $a_{\nu}=b_{\nu}$ at values $0.1,1, 10$. For each such choice, we vary $a_{\tau_\theta} = b_{\tau_\theta} = a_{\tau_\beta} = b_{\tau_\beta}$ at values $0.1,1,10$. For each hyper-parameter combination, MSPE, coverage of the 95\% predictive interval, and the length of the 95\% predictive intervals averaged over all image voxels and network edges are shown. We also compute the Adjusted Rand Index (ARI) between the partitions of influential and uninfluential node indices obtained from a specific combination of hyper-parameters and from the choice $a_\nu = b_\nu = 1$ and  $a_{\tau_\theta} = b_{\tau_\theta} = a_{\tau_\beta} = b_{\tau_\beta} = 1$. All these metrics in Table~\ref{tab_sen_ana} indicate that BMRR is largely insensitive to the choice of hyper-parameters.

\begin{table}[!ht]
\centering
\scriptsize{
\caption{Influential ROIs identified by BMRR in the PPA data analysis along with their posterior probabilities of being influential. See \cite{brainnetome} for Brainnetome atlas ROI naming conventions.}
\label{tab_sel_prob_bmrr}
\begin{tabular}{lclclclc}
\toprule
        \it ROI    & \it Prob. &  \it ROI    & \it Prob. &  \it ROI    & \it Prob. & \it ROI          & \it Prob. \\
\midrule
        BG\_L\_6\_1   & 0.930 & IPL\_L\_6\_2 & 0.722 & MFG\_R\_7\_3  & 0.882 & SFG\_L\_7\_4 & 1.000 \\ 
        BG\_L\_6\_2   & 1.000 & IPL\_L\_6\_3 & 0.800 & MFG\_R\_7\_4  & 0.586 & SFG\_L\_7\_7 & 0.858 \\ 
        BG\_L\_6\_4   & 1.000 & IPL\_L\_6\_5 & 1.000 & MFG\_R\_7\_5  & 1.000 & SFG\_R\_7\_2 & 0.784 \\ 
        BG\_L\_6\_5   & 0.820 & IPL\_R\_6\_5 & 1.000 & MFG\_R\_7\_6  & 0.800 & SFG\_R\_7\_3 & 0.908 \\ 
        BG\_R\_6\_1   & 0.822 & MFG\_L\_7\_1 & 0.964 & MFG\_R\_7\_7  & 1.000 & SFG\_R\_7\_4 & 0.948 \\ 
        BG\_R\_6\_2   & 1.000 & MFG\_L\_7\_2 & 0.596 & MTG\_L\_4\_1  & 0.650 & SPL\_L\_5\_2 & 1.000 \\ 
        BG\_R\_6\_4   & 1.000 & MFG\_L\_7\_3 & 1.000 & PCun\_R\_4\_1 & 1.000 & SPL\_L\_5\_5 & 1.000 \\ 
        CG\_R\_7\_3   & 0.986 & MFG\_L\_7\_4 & 1.000 & PCun\_R\_4\_2 & 1.000 & SPL\_R\_5\_2 & 0.952 \\ 
        CG\_R\_7\_5   & 1.000 & MFG\_L\_7\_5 & 0.526 & PhG\_L\_6\_4  & 1.000 & SPL\_R\_5\_3 & 0.766 \\ 
        IFG\_L\_6\_3  & 1.000 & MFG\_L\_7\_6 & 1.000 & PhG\_R\_6\_5  & 0.726 & Tha\_L\_8\_2 & 1.000 \\ 
        IFG\_R\_6\_2  & 1.000 & MFG\_R\_7\_1 & 0.876 & PoG\_L\_4\_3  & 1.000 & ~ & ~ \\ 
        IFG\_R\_6\_3  & 0.952 & MFG\_R\_7\_2 & 0.910 & PrG\_L\_6\_2  & 1.000 \\
\bottomrule
\end{tabular}
}
\end{table}

\begin{table}[!ht]
\centering
\scriptsize{
\caption{Posterior Predictive Loss Criterion (PPLC) as in as in \cite{gelfand1998model} for values of for BMRR, JR and Horseshoe. $PPLC = G + P$, where G is associated with goodness of fit and P is associated with a penalty related to model complexity.}
\label{tab_pplc}
\begin{tabular}{lccc}
\toprule
Method    & $G$      & $P$     & PPLC \\
\midrule
BMRR      & 2094.40 & 54490.50 & 56584.90 \\
JR       & 2243.31 & 54480.63 & 56723.94 \\
Horseshoe & 2316.32 & 54481.55 & 56797.87 \\
\bottomrule
\end{tabular}
}
\end{table}

\begin{table}[!ht]

\begin{center}
\scriptsize{
\caption{\enspace Sensitivity analysis for the hyper-parameters. Performance measures with different combinations of hyper-parameters, and ARI and proportion of matches of regions selected and unselected with respect to the hyper-parameter combination $a_\nu = b_\nu = 1 = a_{\tau_\theta} = b_{\tau_\theta} = a_{\tau_\beta} = b_{\tau_\beta} = 1$.}\label{tab_sen_ana}
    \begin{tabular*}{\hsize}{@{\extracolsep{\fill}}lccccccccc}
            \toprule
            & \multicolumn{3}{c}{  $a_\nu = b_\nu = 0.1$} &
                \multicolumn{3}{c}{$a_\nu = b_\nu = 1$}   &
                \multicolumn{3}{c}{$a_\nu = b_\nu = 10$}  \\
            \midrule
            \it $a_{\tau_\theta} = b_{\tau_\theta} = a_{\tau_\beta} = b_{\tau_\beta} =$ &
              0.1 & 1 & 10 & 0.1 & 1 & 10 & 0.1 & 1 & 10 \\
            \midrule
            ARI                          & 1     & 1     & 1     & 1     & 1     & 1     & 1     & 1     & 0.981 \\
            Proportion of Matches  & 1     & 1     & 1     & 1     & 1     & 1     & 1     & 1     & 0.996 \\
            MSPE              & 0.823 & 0.819 & 0.825 & 0.825 & 0.823 & 0.825 & 0.824 & 0.821 & 0.826 \\
	    Interval Coverage & 0.914 & 0.918 & 0.906 & 0.908 & 0.908 & 0.906 & 0.912 & 0.914 & 0.908 \\
            Interval Length   & 0.015 & 0.014 & 0.014 & 0.015 & 0.014 & 0.014 & 0.015 & 0.015 & 0.014 \\
            \bottomrule
        \end{tabular*}
        }
    \end{center}
\end{table}

\section{Conclusion and Future Work}\label{Conclusion}
Motivated by multi-modal imaging applications constituting the structural and functional imaging data on patients with PPA, this article develops a regression approach with structural and network-valued objects on a scalar predictor. A novel prior structure is developed jointly on coefficients corresponding to different objects which can simultaneously exploit object characteristics and the linked information between the objects to draw inference on network nodes significantly related to the scalar predictor with uncertainty. The proposed approach is arguably the first statistical multi-modal response regression approach that allows for the flexibility of drawing inference at the ROI level and the voxel level simultaneously and is equipped with identifying brain regions significantly related to the speech rate measuring neuro-degeneration due to PPA with uncertainty. The analysis of PPA data leads to an important understanding of neuro-degeneration patterns for PPA.

The study has some important limitations and considerations. Our results leverage the organizing hierarchy of the Brainnetome atlas and its associated ROIs to connect structural and network information. Thus, conclusions might differ had an alternate brain atlas been used. We emphasize that our proposed technique and its application to the PPA data is based on strong prior neuroanatomical knowledge that the Brainnetome atlas and subsequent parcellation is appropriate for the clinical question of interest. 
In settings where investigators are unsure of the appropriate atlas, greater emphasis should be placed on assessing the sensitivity of results to the selection of alternate parcellations. In addition, PPA being an infrequently studied disease, our results are based on a modest sample size which requires cautious interpretation of results as discussed in Section 2. As an immediate future work, we will extend our approach to incorporate the spatial correlation in the GM image.

\section*{Acknowledgments}
Rajarshi Guhaniyogi acknowledges funding from National Science Foundation Grant DMS-2220840 and DMS-2210672; and National Institute of Health Grant 1R01NS131604-01A1. Aaron Scheffler acknowledges funding from the National Science Foundation Grant
DMS-2210206 and National Institute of Health Grant 1R01NS131604-01A1. We thank the Memory and Aging Center at the University of California, San Francisco for providing the MR images and the patients and their families for the time and effort they dedicated to the research. The study was supported by Grants from the National Institutes of Health (NINDS R01NS050915, NIDCD K24DC015544,  NIA P50AG023501, NIH RF1NS050915)

\bibliographystyle{natbib}
\bibliography{reference}
\end{document}